\documentclass[superscriptaddress,pre,reprint,showpacs]{revtex4-1}

\usepackage{amsmath,empheq, array, bigints}
\usepackage{bm}
\usepackage{amsfonts}
\usepackage{amsthm}
\usepackage{amssymb}
\usepackage{graphicx}
\usepackage{hyperref}
\usepackage{float}
\usepackage{color}

\definecolor{bluecolor}{rgb}{0,0.,1.}

\definecolor{redcolor}{rgb}{.7,0.,0.}

\newcommand{\Exp}[1]{\mathbb{E}\left[  #1 \right]}

\newcommand*\vx{x}
\newcommand*\vh{h}

\newcommand*\tstar{t_\star}

\newcommand*\deltax{\delta_x(\vx)}
\newcommand*\vdelta{\vec{\delta}}

\newcommand*\vddelta{\hat{\delta}}

\newcommand*\p{\pi}

\usepackage{soul}

\newcommand{\ncd}{\newcommand}
\ncd{\mrm}    {\mathrm}
\ncd{\beq} {\begin{equation}}
  \ncd{\eeq} {\end{equation}}

\graphicspath{{./figures/}}

\begin{document}

       \title{Taming chaos to sample rare events:  the effect of weak chaos}

	\date{\today}

	\author{Jorge C. Leit\~ao}
        \affiliation{Max Planck Institute for the Physics of Complex Systems, 01187 Dresden, Germany}
        
	\author{Jo\~ao M. V. P. Lopes}
    \affiliation{Centro de F\'isica das Universidades do Minho e Porto and
Departamento de F\'isica e Astronomia, Faculdade de Ci\^encias,
Universidade do Porto, 4169-007 Porto, Portugal}

	\author{Eduardo G. Altmann}
        \email{eduardo.altmann@sydney.edu.au}
        \affiliation{School of Mathematics and Statistics, University of Sydney, 2006, NSW, Sydney, Australia}
        \affiliation{Max Planck Institute for the Physics of Complex Systems, 01187 Dresden, Germany}

	\begin{abstract}
Rare events in  non-linear dynamical systems are difficult to
sample because of the  sensitivity to perturbations of initial conditions and of complex
landscapes  in phase space. Here we discuss strategies to control these difficulties and
succeed in obtainining an efficient sampling within a Metropolis-Hastings Monte Carlo framework. After reviewing previous
successes in the case of strongly chaotic systems, we discuss the case of weakly chaotic
systems. We  show how different types of non-hyperbolicities limit the efficiency of 
previously designed sampling methods and we discuss strategies how to account for them. We focus on paradigmatic
low-dimensional chaotic systems such as the logistic map, the Pomeau-Maneville map, and
area-preserving maps with mixed phase space.

\end{abstract}

\maketitle

\noindent{\bf
Forecast in chaotic dynamical systems requires the evolution of an ensemble of
trajectories, which  can quickly lead to very different outcomes. If the choice of the ensemble is
compatible with our knowledge of the current  state of the system, we can associate the
probability of an event with the fraction of initial conditions for which it
occurs. Computationally, this strategy is efficient to determine the most likely events
but it struggles to compute rare events, e.g., those at the tail of the distribution of an observable
of interest.  The importance of such {\it extreme events} is that they often cause the
largest impact  and, due to the chaoticity of the system, they can not be easily
  anticipated.  The development of efficient sampling methods, as aimed in this
  paper,  is crucial in these cases because they are able not only to find trajectories
  leading to extreme events but also to estimate their probability (in the original ensemble).
  }

\section{Introduction}

We are interested in 
sampling rare events in chaotic dynamical systems, a problem that has been the subject of
different approaches in the recent
years~\cite{dellago2002transition,tailleur2007probing,geiger2010identifying,iba2014multicanonical,wouters2016rare}.
Given an initial 
condition $x$ in a $d$-dimensional phase space~$\Omega$, $x \in \Omega \subset
\mathbb{R}^d$, the dynamical system $F$ evolves it $x(t) = F^t(x(0))$ until (at time $t_o$) an observable $E_x=E(x(t_o)) \in \mathbb{R} $ is measured. A rare event corresponds 
to an observable $E$ at the tail of the distribution $P(E)$ obtained from an ensemble of
initial conditions selected in  $\Gamma \subset \Omega$  according to a probability measure $\mu$
such that $\int_\Gamma d\mu = 1$, e.g.,  $\mu$ can be simply the phase space volume
(uniform distribution) or  the natural measure of the
dynamical system. Since the dynamics is deterministic, the variability of the events $E$
and the need for its statistical description are solely due to the sensitive to variations
in the initial conditions and not due to an intrinsic random dynamics. 

The importance-sampling method~\cite{bucklew2013introduction} we construct in this paper samples initial conditions
 $x\in\Gamma$ with probabilities different from $\mu$ in order to obtain more samples at
the tail of $P(E)$. However, our goal is still to be able estimate the probability of the
event $P(E)$ of our original
problem (original ensemble of initial conditions).  The strict determinism of the
(chaotic) dynamics is crucial for the design of efficient sampling methods. On the one
hand, the lack of intrinsic randomness poses difficulties to traditional methods, e.g., it
is not possible to distinguish between trajectories based on different noise realizations
and therefore cloning of trajectories requires more refined procedures~\cite{tailleur2007probing,wouters2016rare}. On the other hand, determinism can be explored in order to efficiently search for trajectories leading to rare events.

The goal of this manuscript is to show how to construct efficient Markov Chain Monte Carlo
methods to sample rare trajectories of chaotic dynamical systems. The key ingredient is to
use information about the chaoticity of the last sampled trajectory to construct a proposal
distribution that efficiently finds a new trajectory of interest.  First we review our
previously proposed ~\citep{leitao2017importance} approach (in Sec.~\ref{sec.2}),
which has been successful in different problems involving strongly chaotic systems~\cite{leitao2017importance,leitao2013monte,gupta,tapias}. We
then focus  (in Sec.~\ref{sec.3}) on  deviations from strong chaos and how they pose challenges for the application of the previously
developed methods.

\section{Chaos and Metropolis-Hastings methods}\label{sec.2}

Sampling methods typically exploit the fact that if a given trajectory of interest is
found -- $x$ with a rare $E_x$ -- this can be  used  to find other
trajectories of interest -- $x'$ with $E_{x'} \approx E_x$. The essential step
for the success of the method is to be able to choose the right proposal distribution
of $x'$ given $x$ -- denoted as $g(x'|x)$ --
that guarantees that $x'$ will likely lead to a ``good'' $E_{x'}$. The most natural choice is
to correlate trajectories $x$ and $x'$ to obtain $E_{x'}$ {\it sufficiently close} to $E_x$.
This is an heuristics often used in Statistical
  Physics, implicit in the choice of minimal/local proposals, e.g., in spin systems
  (single spin flip)~\cite{Lopes.phd2006}, in random networks (single link exchange)~\cite{Fischer2015}, and
  in proteins~\cite{Grassberger1997}.  Since our phase space is continuous it is not clear
  what a minimal local proposal would be in our case. In this sense,  the challenge we
  address in this paper is to formalize what {\it sufficiently close} means and to
  construct an efficient proposal $g(x'|x)$ that  achieves it.

The general ideas sketched above are valid for broad classes of sampling methods, but here
we focus on a Metropolis-Hastings setting~\cite{NewmanBarkemaBook,RobertCasellaBook}, in
line with our previous works revised in~\cite{leitao2017importance}. In this setting, starting from $x \in \Gamma$ a new
state $x' \in \Gamma$ is proposed according to $g(x'|x)$. This move can be accepted --
 the new trajectory $x'$ is sampled and the procedure is  repeated from $x'$ -- or
 rejected -- the trajectory $x$ is sampled again and the procedure is repeated from $x$
 with an independent sample from $g(x'|x)$.  This procedure is repeated $n$ times, leading
 to $n$ (correlated) samples. If
 the proposal is ergodic -- all  $x\in \Gamma$ have a non-zero probability to be  sampled for $n\rightarrow \infty$ -- and the acceptance given by
\begin{equation}
a(x'|x) = \min \left( 1, \frac{g(x|x')}{g(x'|x)}   \frac{\p(x')}{\p(x)}    \right) \ \ ,
\label{eq:acceptance}
\end{equation}
the sampled trajectories $x'$ will approach $\p(x)$ for $n\rightarrow \infty$~\cite{RobertCasellaBook}.  The sampling
distribution $\p(x)$ can be chosen at will.  A popular choice is the canonical distribution~\cite{NewmanBarkemaBook}
\begin{equation}
\p(x) = \p(E_x) \propto e^{-\beta E_x} \ \ ,
\label{eq:canonical_ensemble}
\end{equation}
where different regions of the distribution $P(E)$ are sampled when the parameter $\beta$
is varied. Another popular choice is the multi-canonical (flat-histogram) distribution\cite{Berg1991}
\begin{equation}~\label{eq.flat}
  \pi(\vx) \propto \frac{1}{P(E_\vx)} \text{  for }  E_x\in[E_{min},E_{max}],
  \end{equation}
which can be computed (in case $P(E)$ is unknown) also through the Wang-Landau
method~\cite{Wang2001}.

The crucial step to implement a Metropolis-Hastings algorithm to sample chaotic
trajectories is the construction of an efficient proposal $g(x'|x)$
distribution.  The following three steps can be
used to achieve this~\cite{leitao2017importance}:

\begin{itemize}

\item[1.] The goal is to bound the acceptance~(\ref{eq:acceptance}) by making $E_x'$ and
  $E_x$ sufficiently close to each other. Assuming 
  \begin{equation}\label{eq.approx}
    g(x|x') \approx g(x'|x),
  \end{equation}
  the key  remaining term in the acceptance is the ratio $\p(E_{x'})/\p(E_x)$ . We can thus
  express our heuristic more formally by fixing the expectation of this ratio over all
  possible $\vx'$
\begin{equation}
\Exp{\frac{\p(E_{\vx'})}{\p(E_{\vx})} | \vx} = \int_\Gamma
\frac{\p(E_{\vx'})}{\p(E_{\vx})} g(\vx'|\vx) d\vx' = a\ \ ,
\label{eq:expectation_x_prime}
\end{equation}
where $0<a \le 1$ is a constant  (ideally, the constant acceptance rate). Since the
proposal achieves a small variation of $E$, $\p(E_{\vx'})$ can be expanded in Taylor
series around $E_{\vx'} = E_{\vx}$ as
\begin{equation}
\frac{\p(E_{\vx'})}{\p(E_{\vx})} = 1 + \frac{d\log \p(E_{\vx})}{dE} (E_{\vx'} - E_{\vx}) \ \ .
\label{eq:taylor_expansion}
\end{equation}
Introducing Eq.~(\ref{eq:taylor_expansion}) in Eq.~(\ref{eq:expectation_x_prime})  we obtain an
explicit condition
\begin{equation}
\Exp{ E_{\vx'} - E_\vx | \vx } = \frac{a - 1}{d\log \p(E_{\vx})/dE} \ \ .
\label{eq:energy_condition}
\end{equation}

\item[2.] The next step is to compute the correlation time $\tstar$ needed for the trajectories to be close to
  each other in order to achieve condition~(\ref{eq:energy_condition}).
We assume that the observables $E$ of interest are built throughout the $t_o$ times step of the trajectory so that trajectories that
remain close (correlated) in the phase space lead to similar observables $E$. The correlation 
time $\tstar$, $0 \le \tstar \le t_o$ is the time the two trajectories remain ``close'' to
each other, i.e., within a distance $\Delta$ that is smaller than the expected distance between two randomly chosen trajectories in $\Omega$. We assume that, in practice, the two trajectories are identical until $\tstar$, i.e. $x(t) = x'(t)$ for $0 \le t \le \tstar$,and independent for
$t>\tstar$, i.e.  $x'(t)$ is  sampled according to $\mu$ for $t> \tstar$. Explicit expressions
  for   $\tstar$ have been derived for the escape time and  finite time Lyapunov exponent
  (see Ref.~\cite{leitao2017importance}) and  for the dispersion in spatially extended
  (diffusive) systems (see Ref.~\cite{tapias}).

\item [3.] Once the $\tstar$ that guarantees condition~(\ref{eq:energy_condition}) is
  known for a given problem (i.e., for a given observable  $E$ and distribution $\p$)  we
  can generate trajectories $x'$ from one of the following two procedures:  shifting the
  trajectory by a time $\tstar$ backward/forward using the dynamics $x' = F_{\pm
    \tstar}(x)$ (shift proposal); or proposing $x'$ on a neighborhood of size $\delta_x$
  around $x$ (local proposal). The choice of $\delta_x$ in a chaotic system is such that
  the trajectories should remain close to each other up to a time $\tstar$ despite the
  exponential divergence of nearby trajectories, and thus
\begin{equation}
\deltax = \Delta e^{-\lambda_{\tstar}(\vx)\tstar(\vx)}\ \ ,
\label{eq:deltax}
\end{equation}
where $\lambda_t(x)$ is the largest finite-time Lyapunov exponent (FTLE) of the trajectory
(initiated in position $x)$ and $\Delta$ is a constant of the order of $|\Gamma|$.
Depending on the problem, $\lambda_{\tstar}$ can be approximated by $\lambda_{t_o}$ or
(more strongly) by $\lambda_{t\rightarrow \infty}$
(the largest Lyapunov exponent of the system).
  
\end{itemize}

The construction proposed above has been successfully implemented to obtain efficient
Monte Carlo methods in different problems~\cite{leitao2017importance}, including
calculations of finite-time Lyapunov exponents in N coupled oscillators (with N up to
$1024$)~\cite{gupta} and the computation of trajectories with high dispersion in diffusive
systems such as the Lorentz gas~\cite{tapias}. Here we focus on violations of the simplifying assumptions and approximations made above. Violations of strong chaos and uniform hyperbolicity are typical in chaotic dynamical systems, and our interest is to investigate how they affect our sampling methods and how our methods can be modified to account for them. We denote by weak chaos the chaotic dynamics observed in systems that violate the simplifying hypothesis of uniform exponential divergence of initial conditions used above, including systems with marginal stable points and Hamiltonian systems with mixed phase space~\cite{OttBook,Zaslavsky2002}.


\section{Effect of weak chaos on the sampling algorithm}\label{sec.3}

In each of the three subsections below we consider a simple dynamical systems, with
increasingly important (generic) weakly-chaotic features, that violate some of the
simplifying assumptions used above.

\subsection{Logistic map}

A crucial step in our derivation above is that the width of the local proposal, $\deltax$ in Eq.~(\ref{eq:deltax}), guarantees that the trajectory starting from $\vx'$ is within $\Delta$ of the trajectory starting from $\vx$ up to time $\tstar$.
This approximation was based on the assumption of exponential divergence of nearby trajectories, using the maximum FTLE~$\lambda_t(x)$. 
The example we consider below shows how this assumption can be violated due to the coexistence of regions with positive and negative FTLE. Consider the logistic map, defined on $\Omega=[0,1]$ by
\begin{equation}\label{eq.logistic}
x_{t+1} = F(x_t) = 4 x_t(1 - x_t) \ \ .
\end{equation}
Any irrational initial condition $x$ ergodically fills $\Omega$ following $d\mu = \frac{1}{\pi \sqrt{x(1-x)}} dx$. The FTLE in this simple system can fluctuate considerably. The Lyapunov exponent is positive, but finite time estimations can be negative because $|dF/dx| = |4 - 8x|<1$ for $3/8 < x < 5/8$. Particularly problematic are trajectories that come close to $x=1/2$, where  $|dF/dx| =0$.  As a consequence,
the distribution of FTLE of this map, $P(\lambda_t)$, has negative values for any finite
$t$~\cite{Prasad1999}, which implies that there is a non-zero measured set where
$\lambda_t(x) < 0$ for all $x$. 
For any $x$ in this set, and any fixed $\Delta$ in Eq.~(\ref{eq:deltax}), increasing
$\tstar$ leads to a proposal with a width larger than 1 (the phase space size).
This implies that, for large $\tstar$, $x'$ is approximately drawn uniformly from $[0,1]$.
In this situation, it is never expected that the two trajectories $x'$ and $x$ are close
within $\Delta$ up to time $\tstar$, violating our initial assumption.

The crucial violation here is related to the fact that the Lyapunov exponent corresponds to a linear
(first-order) term that correctly describes the divergence of two trajectories in time  in
the limit that their initial separation goes to 0.  The growth of the divergence of two trajectories separated by a \emph{finite distance} $\deltax$ is not necessarily well described by $\exp(-\lambda_t(x) t)$ (see Ref.~\cite{vulpiani2013} for more on finite-size
Lyapunov exponents). This violation is more evident in trajectories $x$ where $\lambda_t(x) < 0$.
To see why, consider a trajectory starting at $x$ on which at time $t_i$, $x_{t_i}\approx
1/2$, and consider that all other $x_t$ are not close to $1/2$.
Since the map is chaotic, up to $t_i$, trajectories starting from $x'$ distanced from $x$ by $\delta_0\exp(-\lambda_{t_i}(x) t_i)$ were approximately, at time $t_i$, within $\delta_0$ of $x_{t_i}$.
However, at that particular point $x_{t_i}$, the derivative is zero and thus the first order approximation predicts that the states $x_{t_i+1}$ and $x_{t_i+1}'$ will be arbitrarily close to each other:
\begin{equation}\label{eq.1sto}
\left|x_{t_i+1}' - x_{t_i+1} \right| \approx \left|\frac{dF}{dx}(x_{t_i})\right| \delta_0 \approx 0 \ \ .
\end{equation}
However, this prediction is might not be accurate because when the first order
term is zero, the second order term is non-zero and dominates:
\begin{equation}\label{eq.2ndo}
x_{t_i+1}' - x_{t_i+1} \approx \frac{1}{2}\left| \frac{d^2F}{dx^2}(x_{t_i}) \right | \delta_0^2 = 4 \delta_0^2 \ \ .
\end{equation}

\begin{figure}[!t]
\centering
\includegraphics[width=0.89\columnwidth]{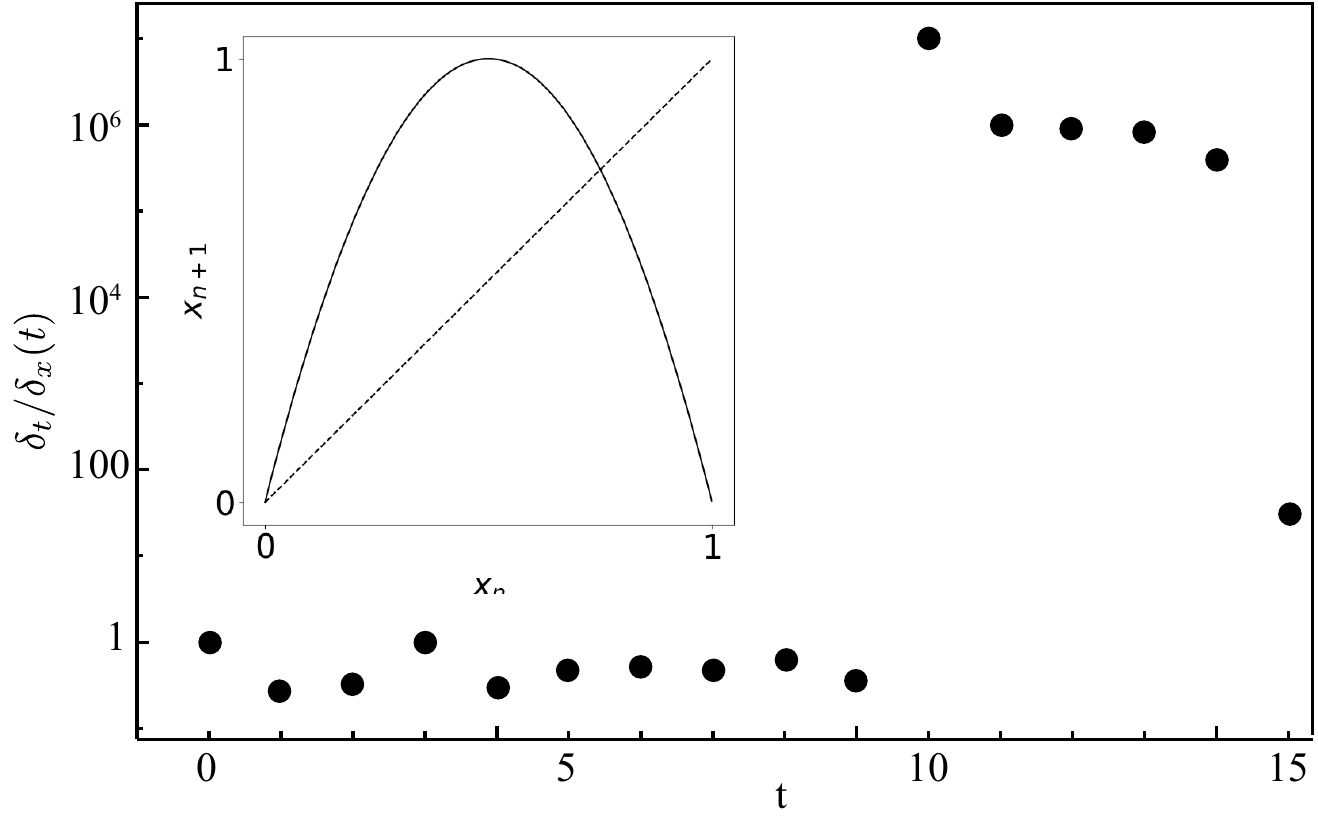}
\caption{
Temporal evolution of the distance $\delta_t \equiv |x_t'-x_t|$ between two close-by initial conditions in the logistic
map~(\ref{eq.logistic}) (see inset).  The relative distance $ |x_t' - x_t|/\delta_x(t)$ (actual
divided by predicted) is shown as a function of time. We first choose one trajectory $x$ with FTLE  $\lambda_{15}(x) \approx -0.15 < 0$
and choose another trajectory $x'=x+10^{-5} \exp(-\lambda_{15}(x) 15)$.
Up to time $t_i = 9$, the actual (numerically obtained) distance $|x_t' - x_t|$ is well described by
$\delta_x(t)=\delta_x(0) \exp(\lambda_t(x) t)$ so that $\delta_t/\delta_x(t) \approx 1$. This dramatically changes at 
$t_{i+1}=10$ because $x(t_i) \approx 0.5$ and  the distance becomes much larger than expected. 
}
\label{fig:logistic_delta_ratio}
\end{figure}

\begin{figure*}[!ht]
  \centering
  \includegraphics[width=1.6\columnwidth]{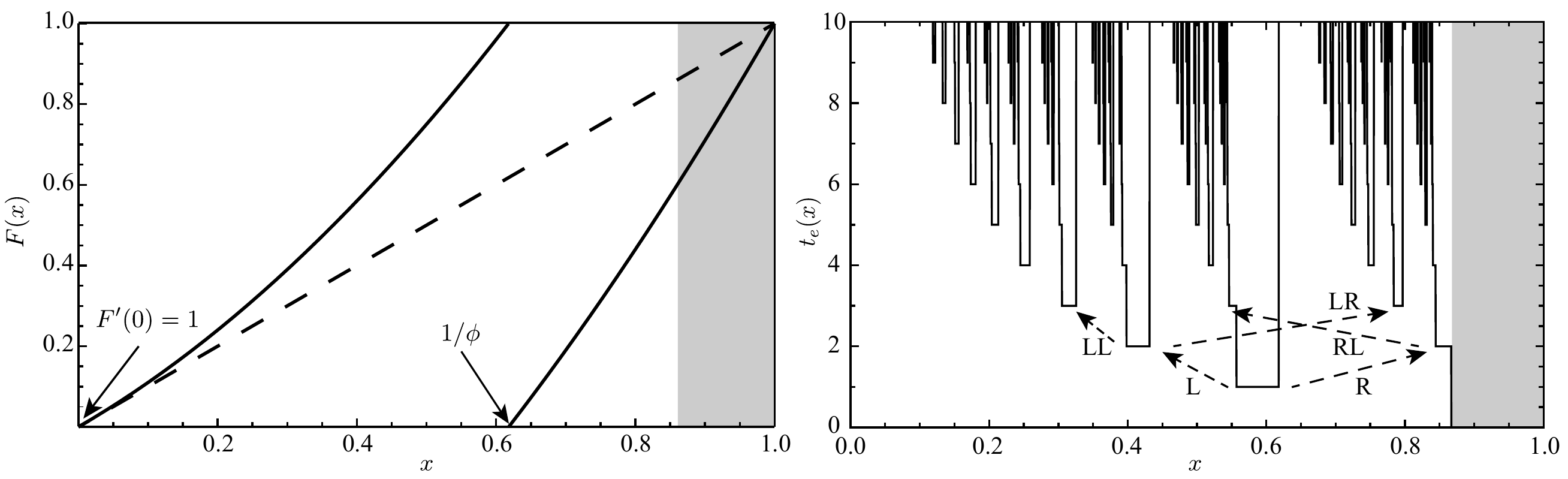}
\caption{
  The open Pomeau-Maneville map. Left: the Pomeau-Maneville map and the exit region $\Lambda$ in gray. Right: the escape time function of the map. The landscape is fractal and the intervals with constant escape time $t_e$ have a symbolic sequence associated to them: each new interval at $t_e+1$ is constructed to the left (L) or to the right (R) of an interval at $t_e$, except when the interval at $t_e$ was to the right from it's own previous $t_e-1$. It is thus a restricted symbolic dynamics $s_{1}s_{2}s_{3}...s_{t}$ with the forbidden sequence RR.
}
\label{fig:maneville_landscape}\label{fig:maneville_escape_time} 
\end{figure*}

We confirmed numerically the appearance of the behaviour described above, which is known
as glytch~\cite{ChaosBook}. We consider two trajectories initially separated by $\Delta
\exp(-\lambda_t(x) t)$, and we compare the distance in time, $\delta_t \equiv |x_t' -
x_t|$, with the distance $\delta_x(t) = \Delta \exp(\lambda_t(x) t)$ expected based on the
first order approximation~(\ref{eq.1sto}).
A violation of the assumption happens when the ratio $r(t) \equiv \delta_t/\delta_x(t)$ is
different from $1$.
Figure~\ref{fig:logistic_delta_ratio} shows a representative example of this simulation,
which confirms that the ratio $r(t)$ can change abruptly, becoming orders of magnitude
different from $1$ (indicating that the distance $|x_t' - x_t|$ is much larger than the expected distance given by $\delta_x(t)$).
This could be fixed  by decreasing the initial distance $\Delta$, but the crucial point here is that $\Delta$ strongly depends on the particular $x$ and can be orders of magnitude different for different $x$ (e.g. one with $\lambda_t > 0$ vs. one with $\lambda_t < 0$).
In other words, the assumption that is violated in the logistic map is that there is a $\Delta$ independent of $x$ that makes $\deltax$ in Eq.~(\ref{eq:deltax}) to guarantee a correlation time $\tstar$ between any two trajectories.
The results above do not imply that Metropolis-Hastings cannot be used in systems where
$\lambda_t < 0$ for some states, they  imply that Eq.~(\ref{eq:deltax}) has to be
extended, e.g.,  to make $\Delta$ dependent on $x$. In the next section we consider a
similar issue, arising when periodic orbits show zero Lyapunov exponents. 

\subsection{Pomeau-Maneville map}
\label{sec:pm}

One important approximation in the derivation of $\tstar(\vx)$ in point 2. of Sec.~\ref{sec.2}
above is that, for $t>\tstar$,  $\vx_{\tstar}'$ is independent of $\vx_{\tstar}$ .
This approximation was based on the notion that trajectories diverge exponentially and
thus two trajectories  are separated by $\Delta \approx 1$ at $t=\tstar$ will rapidly become independent of each other. This approximation is
naturally violated when the (local) divergence of nearby trajectories is not exponential, and our goal here is to explore the
consequences of this violation to our sampling method.

Let us analyze one simple one-dimensional system where non-exponential divergence is present, the Pomeau-Maneville map defined in $\Omega=\Gamma=[0,1]$ by
\begin{equation}
x_{t+1} = F(x_t) = x_t + x_t^2 \mod 1 \ \ .
\label{eq:pm_map}
\end{equation}
This map is a model for intermittency, a phenomenon on which trajectories irregularly alternate between regular and chaotic motion~\cite{OttBook}.
The intermittency in this system appears because $\frac{dF}{dx}=1 + 2x$ and thus the fixed
point $x=0 = F(0)$ is a non-hyperbolic point. The ergodic invariant measure $\mu$ is non-normalizable as it diverges at the fixed point as $d\mu \sim \frac{1}{x} dx$~\cite{Thaller1995}. More general Pomeau-Maneville maps consider a generic power $z$ instead of $2$ in Eq.~(\ref{eq:pm_map})~\cite{Niemann}. From this point of view, the case treated here ($z=2$) is special because it lies at the border between normalizable and non-normalizable $\mu$ and it would be interesting to generalize our results to $z \neq 2$.

The observable we are interested in is the time a trajectory takes to escape an open Pomeau-Maneville
map, achieved leaking~\cite{Altmann2013} the map by adding an exit region $\Lambda$ so that trajectories $x \in \Lambda$ are removed ($t_e=0$). Choosing $\Lambda = [\ell, 1]$ for the map~(\ref{eq:pm_map}), with
\begin{equation}
\ell=\frac{1}{2}\left(-1+\sqrt{3+2\sqrt{5}}\right) \ \,,
\label{eq:leaks}
\end{equation}
ensures that the function relating the escape time $t_e$ to
the initial condition $x$ can be described by a symbolic sequence with forbidden
sequences, as shown in Fig.~\ref{fig:maneville_landscape}.
The distribution of escape times $P(t_e)$ is known to have a power-law tail $P(t_e)
\sim t_e^{-\alpha}$ with an exponent $\alpha = 2$. This implies that the average escape time $\langle t_e \rangle$ diverges, a strong form of intermittency (or stickiness at the origin). Since the escape time varies over orders of magnitude, it is
natural to consider as an observable the logarithm of the escape time $E_\vx = \log
t_e(\vx)$.
The distribution of $E_\vx = \log t_e(\vx)$ is then exponential with an
exponent $\alpha' = \alpha + 1 = -1$. Our interest is to estimate $P(t_e)$ and sample
trajectories at the tail of this distribution.

\begin{figure}[!ht]
  \centering
  \includegraphics[width=0.78\columnwidth]{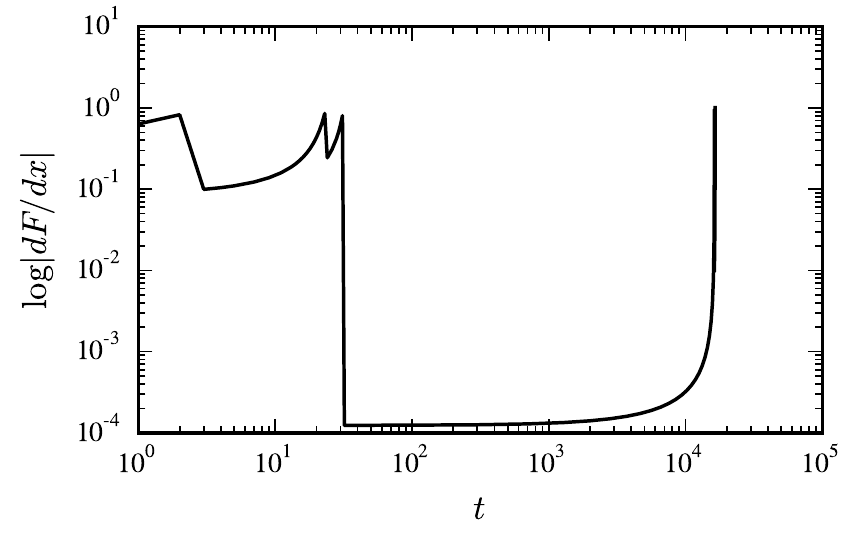}\\
  \includegraphics[width=0.78\columnwidth]{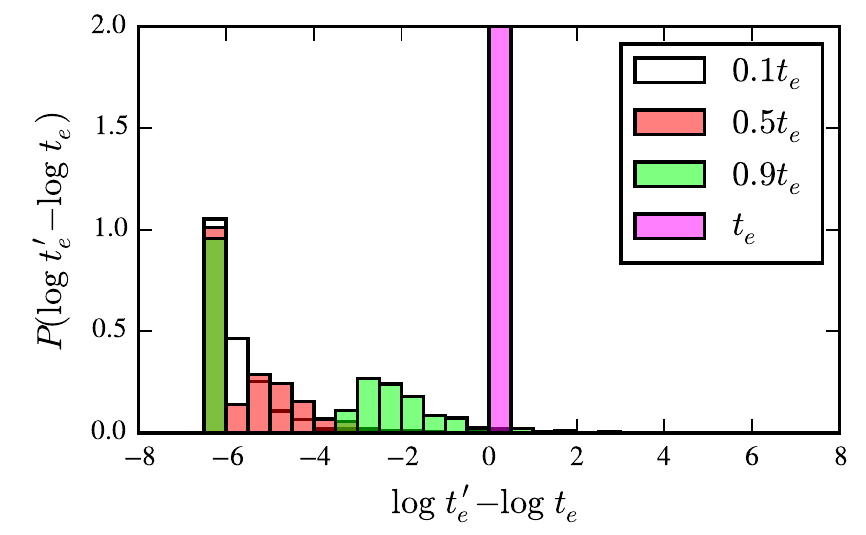}\\
      \includegraphics[width=0.78\columnwidth]{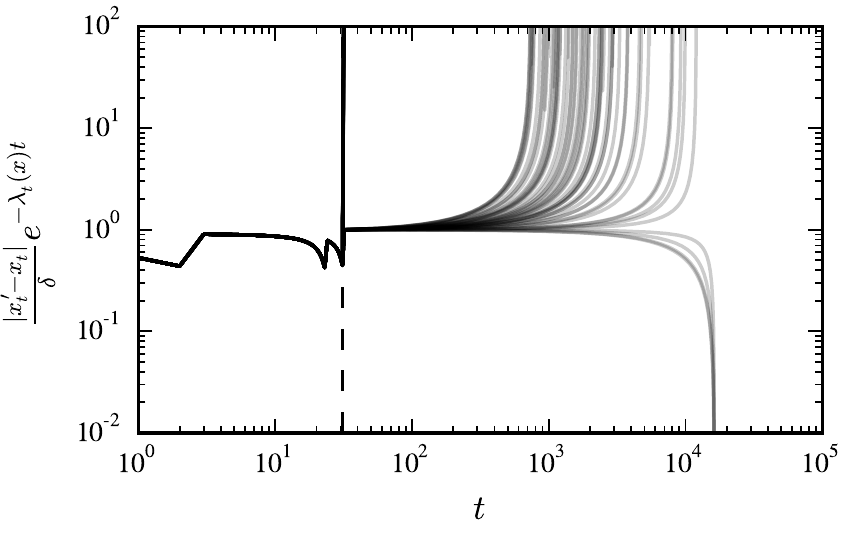}\\
      \caption{
          Searching for local proposals in the open Pomeau-Maneville map~(\ref{eq:pm_map}).
          (Top)  Representation of $|\log d F/dx |$ of a typical long-living trajectory: it
          starts in the chaotic region (high derivative), at $t_i=31$ it is injected very
          close to $x=0$ (low derivative),   it then shows very slow divergence until it leaves this region and eventually   hits the exit region and  escapes at $t=16458$.
The FTLE of this trajectory is the arithmetic mean of this curve, which shows large fluctuations as a function of time due to the intermittency in the trajectory.
(Middle) $P(\log t_e(x') - \log t_e(x)|x)$, where $x$ is the trajectory represented in the top panel
($t_e(x) = 16458$) and $1000$ different $x'$'s are generate according to $\deltax$ in
Eq.~(\ref{eq:test_delta}), for $\Delta=1$ and different values of $\tstar$ (see legend).
(Bottom) Individual trajectories starting at $x'$ (generated for $\tstar=0.9 t_e$) nearby from $x$ are shown as a thin black
lines. Approximately half of them escape a the time $t_i =31$, when $x$ is injected to the
non-hyperbolic point $0$ (see top panel).
}
\label{fig:pm_trajectory}
\label{fig:maneville_conditional_trajs}
\label{fig:maneville_conditional_dist}
\end{figure}

Qualitatively, a typical long living trajectory can be pictured by a trajectory that, for a time $t_{chaos}$, behaves as if it was a chaotic trajectory, and that at some time, denoted here as a time $t_i$, is injected close to the non-hyperbolic point $x=0$.
The trajectory then spends a long time $t_{stick}$ close to $0$, until it eventually leaves the region, returning to a chaotic movement.
An example of such a trajectory is shown in Fig.~\ref{fig:pm_trajectory}a.

We now investigate how the recipe from Sec.~\ref{sec.2} can be used to construct an efficient proposal
distribution for the Pomeau-Maneville map, focusing on the main differences between this map and other strongly chaotic system for which the recipe has worked in the past.
The two major differences here are: a) the observable is $E = \log t_e$, instead of $t_e$; and b) there is a non-hyperbolic fixed point at $x=0$.
We obtain numerical insights on this problem by starting from a trajectory $x$ with a
high escape time $t_e(x) $ and searching for different $x'$,  obtained
adding to $x$ a small perturbation of typical size $\delta_x$ (see
Appendix~\ref{sec.halfgaussian} for details) given by
\begin{equation}
\delta_x(\vx,\tstar) = \Delta e^{-\lambda_{\tstar} \tstar} \ \ .
\label{eq:test_delta}
\end{equation}
This equation is similar to Eq.~(\ref{eq:deltax}), but here instead of using a theoretically
derived $\tstar$ we use it as a free-parameter that defines a scale. Using this proposal,
we measure $P(\log t_e' - \log t_e|\vx)$ for different $\tstar$.
The goal is to test the hypothesis that proposing with Eq.~(\ref{eq:deltax}) guarantees that
the two trajectories remain, on average, close together up to time $\tstar$.  If the
assumption holds in this system, choosing $\tstar = q t_e$, with $0<q<1$, would imply that on average
$t_e(\vx') \gtrapprox \tstar = q t_e(\vx)$. 
In terms of the logarithm, this would imply that 
\begin{equation}
\log t_e(\vx') \gtrapprox \log t_e(\vx) + \log q\\
\label{eq:test_eq}
\end{equation}

The outcome of a numerical experiment that implements the ideas above is shown in
Fig.~\ref{fig:maneville_conditional_dist}.
We focus on a  trajectory $x$ with an escape time $t_e(x) = 16458$, or $\log t_e \approx
10$ (Fig.~\ref{fig:maneville_conditional_dist}, upper panel).   The variation in the observable $P(\log t_e' - \log t_e | x, \tstar)$ for different values of
$\tstar$ shows (Fig.~\ref{fig:maneville_conditional_dist}, middle panel) that, independently of $\tstar$,  most trajectories show a
$\log t_e' - \log t_e$ much smaller than the expected value from Eq.~(\ref{eq:test_eq}) (e.g.,
for $q=0.5$, $\tstar = 0.5 t_e$, we would expect $\log t_e(\vx') -\log t_e(\vx) \gtrapprox
- 0.7$).  In fact, independent of $\tstar$, almost 50\% of all trajectories shows $\log
t_e(\vx') -\log t_e(\vx) \approx -6.5$.
This shows that the assumption that Eq.~(\ref{eq:deltax}) guarantees that the states are
close up to $\tstar$ is violated here due to the non-hyperbolic nature of the point $x=0$.
Still, our result does indicate a dependence of $\log t_e(\vx') - \log t_e(\vx)$ on
$\tstar$, which suggests that it may still be possible to derive a distance between $\vx'$
and $\vx$ that leads to a bounded acceptance.
To investigate this possibility, we repeat the approach done previously for the logistic
map (in Fig.~\ref{fig:logistic_delta_ratio})
and plot
(in Fig.~\ref{fig:maneville_conditional_dist}, lower panel)   the expected divergence
given by the first order term of the Taylor expansion with the actual distance between the
trajectories, for different trajectories $x'$ generated with $\tstar = 0.9 t_e$.
We find that there are many  trajectories that largely deviate from $x$ at the time $t_i$
when the trajectory is injected close to the critical point $x\approx 0$. Half of the
trajectories quickly exit the system (being responsible for the high peak around $\log
t_e(\vx') -\log t_e(\vx) \approx -6.5$), they  correspond to points that in Fig.~\ref{fig:maneville_conditional_dist} have a low
escape time $t_e(x')$.

The numerical observations reported above can be understood analytically by focusing on  the
injection of the chaotic trajectory into the trapping point $x=0$. This injection happens around the
pre-image of $x=0$, which is $x=1/\phi$, where $\phi$ is the golden ratio.
At $1/\phi$, the map $F$ in Eq.~(\ref{eq:pm_map}) is discontinuous.
For a trajectory to be long living, it must approximate $0$, which requires its pre-image
to be very close but larger than $1/\phi$, i.e.,  $x=1/\phi + \epsilon$ (with $\epsilon>0$).
The proposal distribution is a normal distribution around $x$, and therefore half of the trajectories can be at $x<1/\phi$.
These will not be mapped close to $0$ and therefore most likely they quickly leave the system.
The orbits mapped $\varepsilon$ close to $0$ evolve initially as
$x_t \approx \varepsilon + t \varepsilon^2$ (in order $\varepsilon^2$). The escape time
$t_e(x)$  is proportional to the time $x$ takes from $\varepsilon$ to $2\varepsilon$
and therefore $t_e \propto 1/\varepsilon$. The linear divergence close to $x\approx 0$
implies also that the total divergence of such trajectory is
$D = \sum_{i=0}^{t_e(\varepsilon)} \log (1+2 x_i) \approx \sum_{i=0}^{t_e(\varepsilon)} 2
x_i  \propto t_e^2$. We performed numerical simulations
that confirm these two scaling and fix the pre-factors as $t_e = 1/\varepsilon$ and $D=3.3 t_e^2$.

We now use the results discussed above  to obtain an expression for $\deltax$
that guarantees a given variation of $\log t_e(x') - \log t_e(x)$ -- as required by
Eq.~(\ref{eq:energy_condition}) -- and that can thus be used to construct an efficient
proposal distribution $g(x'|x)$.  Let $\varepsilon(x)$ be the smallest distance of $x_t$
from 0 for all $t = 1,...,t_e(x)$, and let $t_i(x)+1$ be the time at which this happens
(i.e., at $t_i(x)$, $x$ is injected close to $x=0$ ). Assuming that there are no
re-injections,  the time from $t_i$ until the state crosses $1/\phi$, $t_e^*(x)$ is the
leading contribution to $t_e(x)$, i.e. $t_e^*(x) \approx t_e(x)$.
 Using that $t_e
=1/\varepsilon$ we can write $\Delta E(x) = E(\varepsilon(x')) - E(\varepsilon(x))$ as
\begin{equation}
\Delta E(x) = \log t_e(\varepsilon') - \log t_e(\varepsilon) = \log \varepsilon - \log \varepsilon' \ \ .
\end{equation}
Introducing $\delta_i(x) \equiv \varepsilon' - \varepsilon$  in the equation above and
solving for it we find
\begin{equation}
\delta_i(x) = \left(e^{-\Delta E_x} - 1\right)\frac{1}{t_e(x)}.
\label{eq:maneville_deltai}
\end{equation}
To guarantee that the two states are within $\delta_i(x)$ at time $t_i$, the distance $\deltax$
between $x'$ and $x$ at time $t=0$ should be (according to Eq.~(\ref{eq:deltax}))
\begin{equation}
\deltax = \Delta e^{-\lambda_{t_i}(x) t_i} \delta_i(x) = \Delta e^{-\lambda_{t_i}(x) t_i} \left(e^{-\Delta E_x} - 1\right)\frac{1}{t_e(x)}
 \ \ .
\label{eq:manneville_deltax_step}
\end{equation}
The result above depends on $t_i(x)$, while we would like to express it as a function of our
observable $t_e(x)$.  After injection the divergence is very small, and thus we replace
$\lambda_{t_i} t_i$ by $\lambda_{t_e} t_e$ but we additionally multiply this expression by
the total  divergence  $D(\varepsilon) \propto t_e(\varepsilon)^2$  discussed above (the
total divergence $D(x, t_e)$ is the product of the divergence up to $t_i$ and the
divergence from $t_i$ to $t_e$). Our final results is then
\begin{equation}
\deltax = \Delta \left(e^{-\Delta E_x} - 1\right) e^{-\lambda_{t_e}(x) t_e} t_e(x) \ \ ,
\label{eq:manneville_deltax}
\end{equation}
which relates the scale that two states $x$ and $x'$ should be in order to achieve a given
expected variation in the observable $E$. This expression replaces Eq.~(\ref{eq:deltax}) for the Pomeau-Maneville map~(\ref{eq:pm_map}) considered here. The essential new feature is the appearance of the multiplying term $t_e(x)$. The term $e^{-\Delta E_x} - 1$, which ideally should incorporate the desired $\Delta E$ computed using Eq.~(\ref{eq:energy_condition}), does not show a strong dependence on $t_e(x)$ so that in practice we considered it to be a constant (i.e., we incorporate it in the proportionality constant $\Delta$).

\begin{figure}[!t]
  \centering
  \includegraphics[width=0.95\columnwidth]{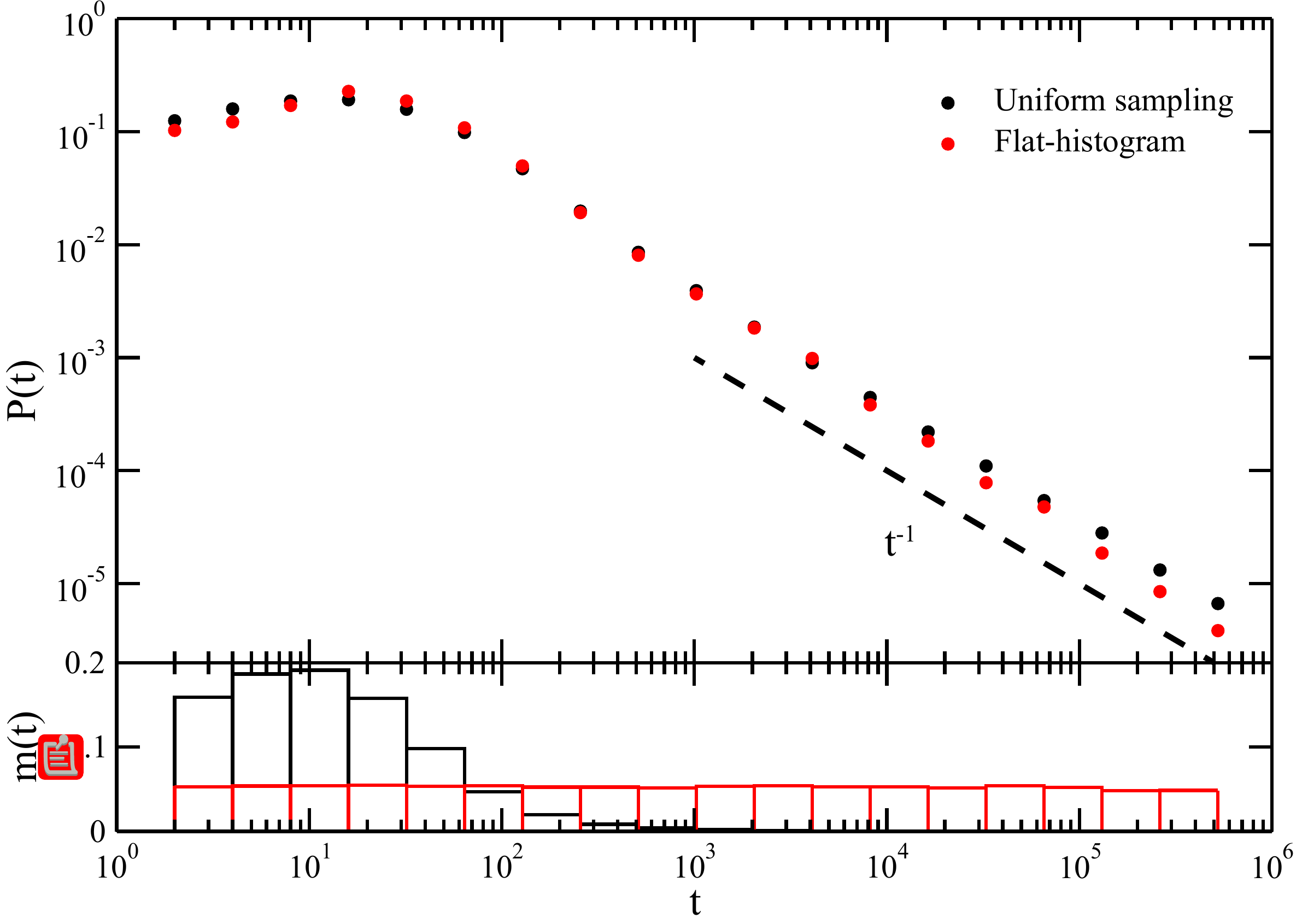}\\
\includegraphics[width=0.95\columnwidth]{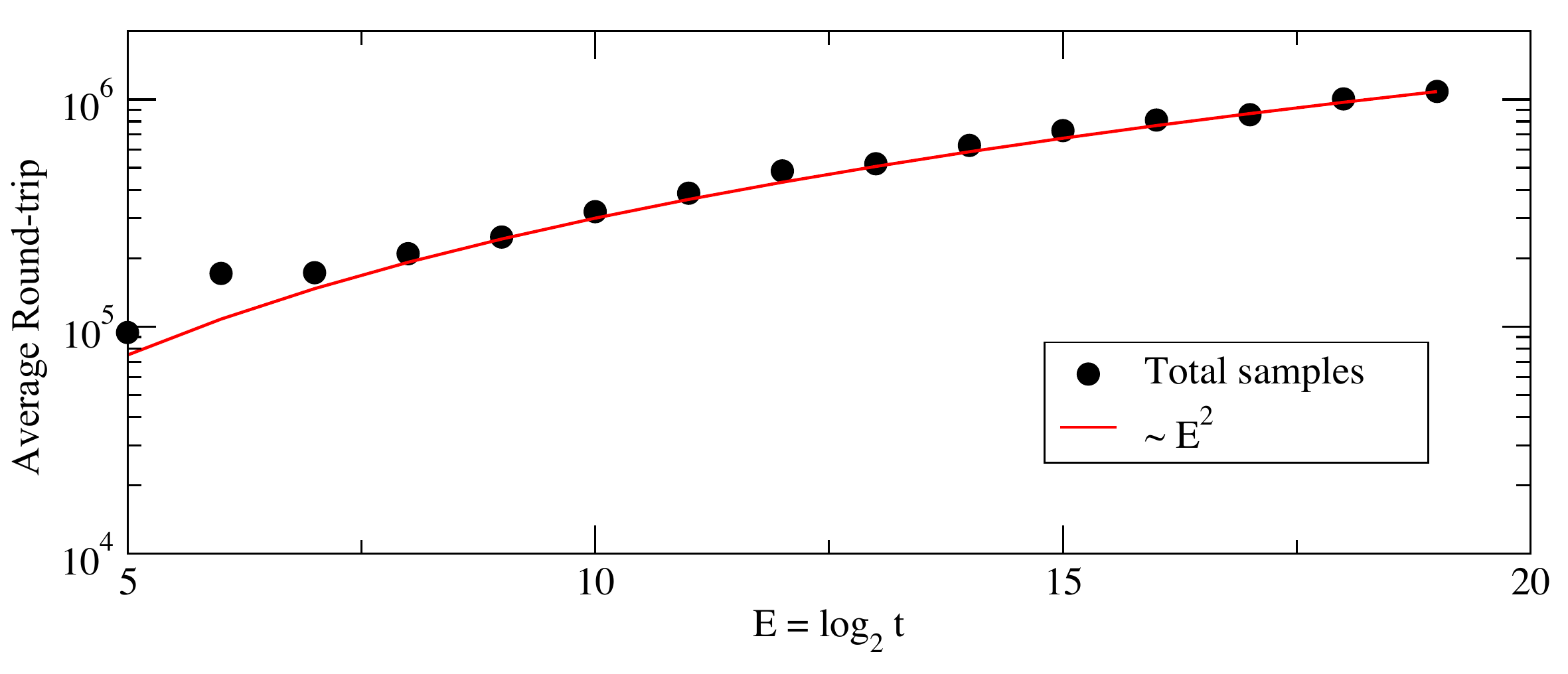}
\caption{ Efficient sampling of trajectories in an intermittent map.
  (Top) The escape time distribution ($P(\log t_e)$) of the Pomeau-Maneville map~(\ref{eq:pm_map}).
 The black symbols were obtained choosing initial conditions uniformly in $\Omega=[0,1]$,
 the red symbols are the results of our importance sampling simulations,  and the dashed
 line is the theoretical asymptotic scaling, $t_e^{-1}$. The x axis is $t_e$ in
 logarithmic scale and the distribution was built using bin size one in  $E = \log_2 t_e$.
 The importance sampling simulation is a Metropolis-Hastings with the Wang-Landau algorithm, using
  the proposal given by Eq.~(\ref{eq:manneville_deltax}). The lower plot shows the histogram $m(t)$ flat in the variable $E = \log_2(t_e)$.
The simulation used 10 Wang-Landau refinement steps, with 5 round-trips on each refinement step. We use Eq.~(\ref{eq:manneville_deltax}) with $\Delta (e^{-\Delta E_x} - 1)=0.1$.
(Bottom)
The average round-trip time $\tau$ of Metropolis-Hastings scales polynomially with maximum
$E$. A round-trip is defined as a movement from $E_{min}=0$ to $E_{max}=E= \log_2 t_e$, for various maximal  $t_e$. 
Each point is the average over 32 round-trips.
}
\label{fig:maneville_confirm}
\label{fig:maneville_efficiency}
\end{figure}

We test the accuracy and usefulness of the results above through numerical simulations
of a flat-histogram~(\ref{eq.flat})  simulation, as reported in Fig.~\ref{fig:maneville_confirm}. The
success on the estimation of $P(t_e)$ can be seen on: (i) the agreement with the
traditional uniform sampling, not only in the tail  of $P(t_e)$ but also for short
times; and (ii) the polynomial scaling of the computational  efficiency  (round-trip
scales as $E^2$ for increasing $E$). This confirms that the proposal
distribution derived in Eq.~(\ref{eq:manneville_deltax}) achieves its goal in obtaining an
efficient Metropolis-Hasting simulation.  While the specific derivation presented here is
valid for the Pomeau-Maneville map only, for which analytical results exist, the reasoning
of deriving a $\deltax$ that allows to change $\log t_e$ can be applied more generally to
maps with marginally unstable points.  The main conclusion here is that the methodology
summarized in Sec.~\ref{sec.2} is applicable to weakly chaotic open systems with power-law
distributions of the escape time, and reinforces the thesis that the strength of this
methodology lies in its ability to construct the proposal distribution from the properties
of the dynamical system under investigation.

\begin{figure}
  \centering
     \includegraphics[width=0.89\columnwidth]{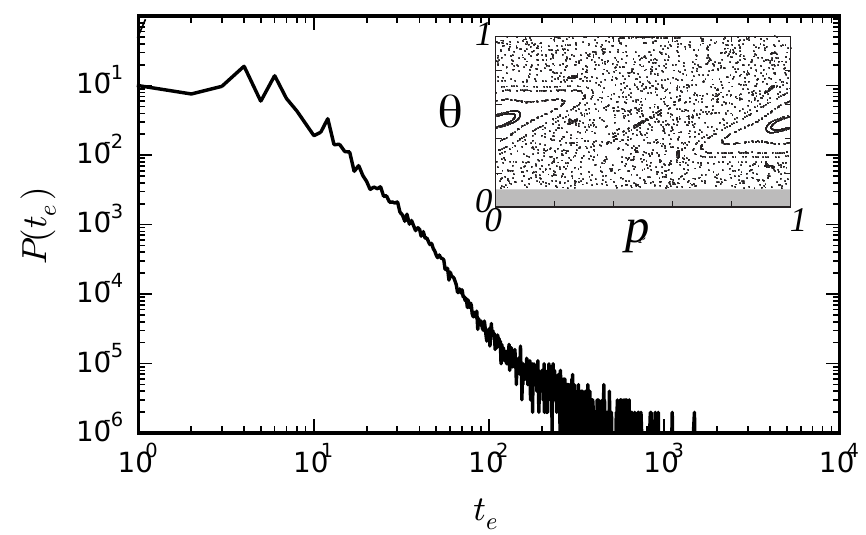}
\caption{
Escape time distribution $P(t_e)$ of area-preserving maps with mixed phase-space systems
shows power-law tails. The system is the standard map~(\ref{eq:standard_map}) with $K=2.1$ and exit region $\Lambda =
[0,1]\times[0,0.1]$, see the Inset  for the iteration of multiple trajectories in the
phase space (notice the KAM island around the elliptic fixed point at $(0,0.5)$). The
escape time distribution $P(t_e)$ was computed by starting $10^6$ initial
conditions uniformly on the first image of the exit region ($F(\Lambda)$).
}
\label{fig:open_standard_map}
\end{figure}

\begin{figure}[!t]
  \centering
 \includegraphics[width=0.86\columnwidth]{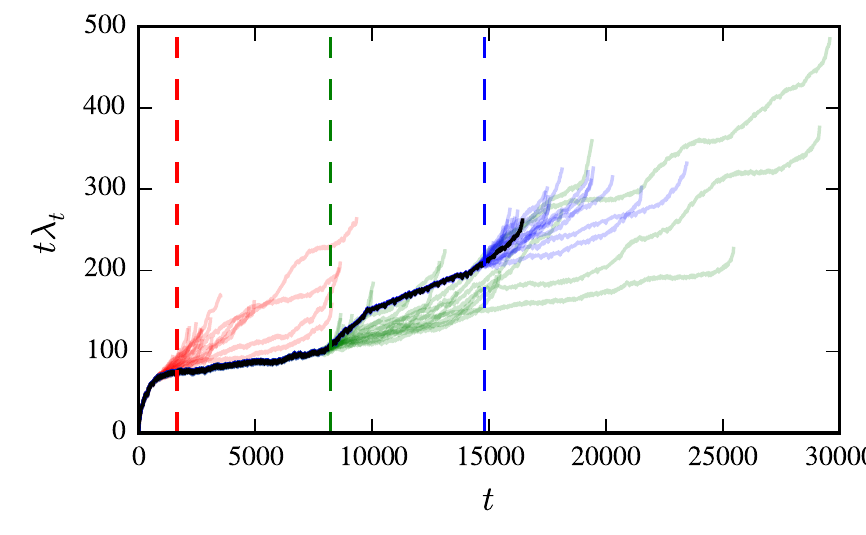}
 \includegraphics[width=0.86\columnwidth]{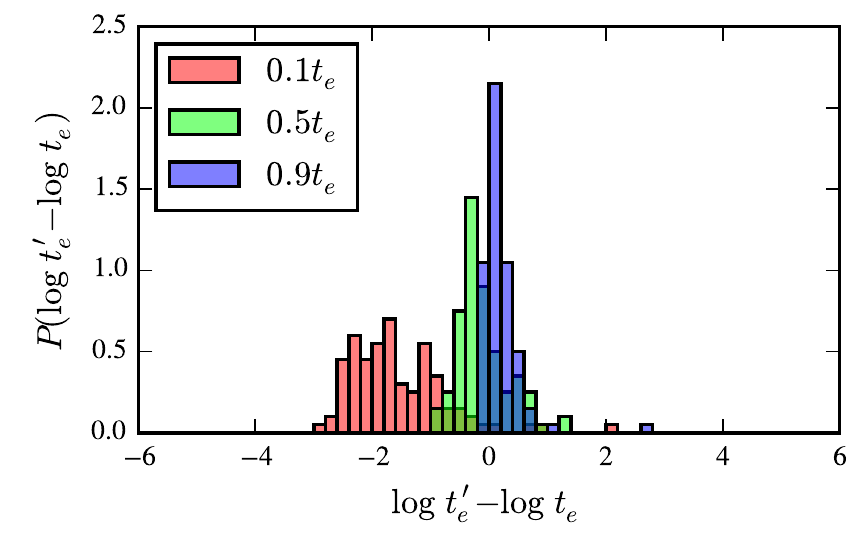}
 \includegraphics[width=0.86\columnwidth]{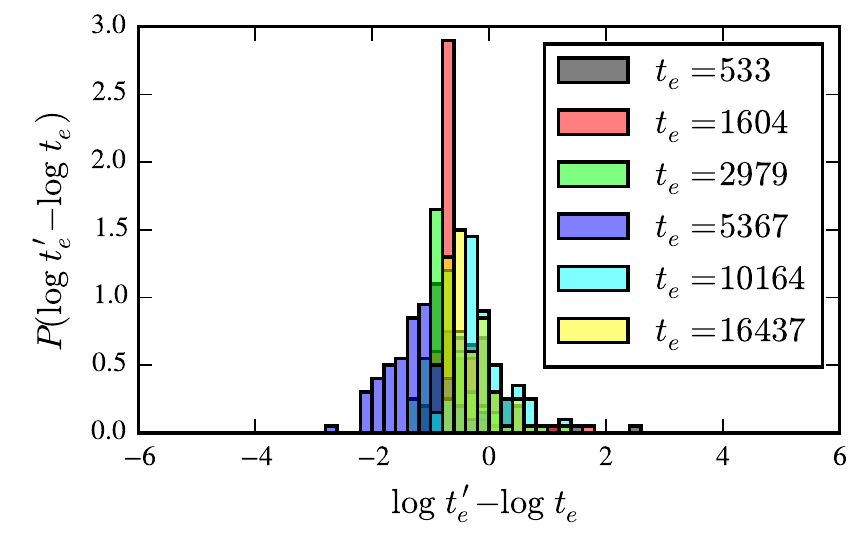}
\caption{Searching for local proposal in the open standard map~(\ref{eq:standard_map}) with
  $K=2.3$. Starting from a randomly selected trajectory $x$ with escape time $t_e$ (in the
  tail of $P(t_e)$),  we  sample nearby trajectories in a neighborhood given by
  $\exp(-\lambda_{\tstar}(\vx) \tstar)$ (Eq.~\ref{eq:test_delta}, $\Delta=1$) for
  different values of $\tstar$ (given as multiples of $t_e$). (Top) Single trajectory $x$
  (in black) and multiple 
  trajectories $x'$ obtained for $\tstar = 0.1 t_e(\vx)$ (red), $\tstar = 0.5 t_e(\vx)$ (green)
  and $\tstar = 0.9 t_e(\vx)$ (blue). Trajectories are plotted until they leave, i.e. the last time corresponds to $t_e(\vx')$.
The vertical dashed lines represent $0.1 t_e(\vx)$, $0.5 t_e(\vx)$ and $0.9 t_e(\vx)$
respectively. (Middle) The conditional probability of $t_e(\vx') = t_e'$ around a particular
state $\vx$ (randomly generated) with an escape time $t_e(\vx)=14995$ ($E=\log_2 t_e
\approx 3$) and different $\tstar$ (see legend).
(Bottom) Histograms for trajectories $x$ with six different values of $t_e(\vx)$ (different colors, see caption) and fixed $\tstar=0.5 t_e$. The distribution of distances $\log t_e(\vx)' - \log t_e(\vx)$ with
$\vx'$ proposed with a correlation time $\tstar(\vx) = 0.5 t_e(\vx)$ from $\vx$ remains
similar with increasing $t_e(\vx)$.
All simulations were made with arbitrary precision~\cite{github}. The
$\lambda_{t_e}(\vx)$ used in Eq.~(\ref{eq:deltax}) was computed by generating a random unitary vector $\vh$ (same as the one used to generate $\vx'$) and evolving it in the tangent space, by multiplying it by the Jacobian matrix $J_t$.
}

\label{fig:standardMap}
\end{figure}

\begin{figure}[!t]
  \centering
 \includegraphics[width=0.8\columnwidth]{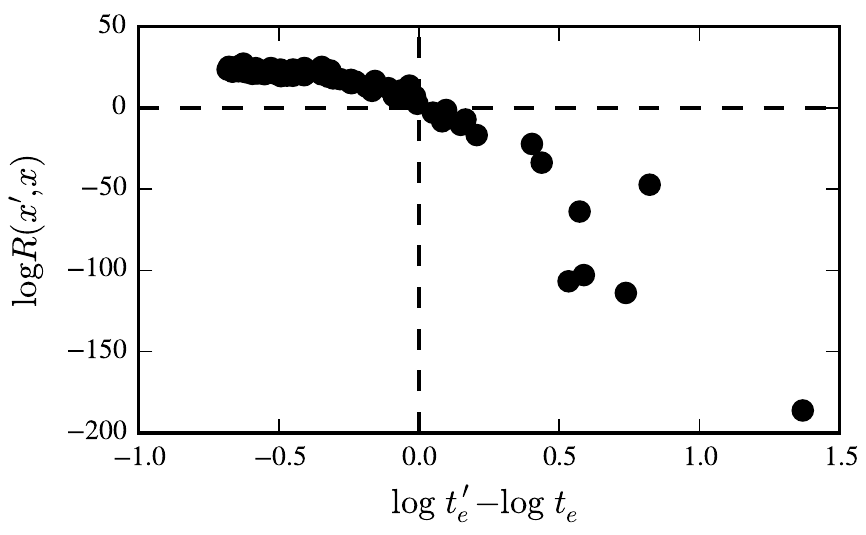}
 \caption{
   Strong variation of the ratio of the proposal distributions, responsible for the low
acceptance rate.  The system is the open standard map with $\tstar(\vx) = t_e(\vx)/2$ and
$t_e=16437 \approx 2^{14}$ (as in Fig. \ref{fig:standardMap}). 
The x-axis shows $\log t_e$ of $t_e(\vx')$ and $t_e(\vx)$.
The y-axis shows $\log R(\vx',\vx)$, where $R(\vx',\vx) = \delta_x(\vx)/\deltax(\vx')$ is
the ratio of the proposal widths $\delta$ at $\vx$ and $\vx'$. This is very different
from $1$ -- contrary to approximation~(\ref{eq.approx})  -- turning the acceptance
rate~(\ref{eq:acceptance}) extremely small.
 }
\label{fig:standardMap3}
\end{figure}

\subsection{Standard Map}
\label{sec:sm_sec}

The previous section considered an example of a weakly chaotic system for which, due to its simplicity, it was possible to derive a relationship between the distance of two trajectories $\vx' - \vx$ and the difference in their respective observables, $E_{\vx'} - E_\vx$.
This motivates us to consider a more challenging problem, which shows a similar type of
survival probability distribution $P(t_e)$ but for which no analytical results
are known.
We consider the area-preserving standard map $(p_{t+1},\theta_{t+1}) =
F(p_t,\theta_t)$ given by~\cite{OttBook}

\begin{equation}
F(p, \theta) = \begin{cases}
p + K/(2\pi) \sin(2\pi \theta) \mod 1\\
\theta + p + K/(2\pi) \sin(2\pi \theta) \mod 1\\
\end{cases} \ \ .
\label{eq:standard_map}
\end{equation}
This map is a paradigmatic example of the KAM scenario of mixed phase-space Hamiltonian
systems~\cite{OttBook,Zaslavsky2002}. For the parameters $K$ we use in our numerical investigations ($K\gtrapprox 2$), chaotic and regular components with non-zero measure
coexist in the phase space. Introducing an exit region in the chaotic component, the escape
time distribution of trajectories started in the same component follow $P(t_e) \sim t_e^{-\alpha}$, as shown in
Fig.~\ref{fig:open_standard_map} (which again motivates the observable $E_\vx = \log
t_e(\vx)$). The intermittency (or stickiness) in the standard map is weaker than the one observed in the Pomeau-Maneville map: there is (normalizable) invariant measure $\mu$ (the phase space area), the exponent is $2 < \alpha < 3$, and the average escape time $\langle t_e \rangle$ exists. The existence of a  universal asymptotic exponent $\alpha$ has been long
conjectured~\cite{Karney,Chirikov} and  the subject of extensive theoretical and
numerical investigations~\cite{Meiss1986,Zaslavsky2002,Cristadoro2008}. The numerical investigations in these works considered uniformly chosen initial conditions and our goal is to investigate whether more efficient choices of initial conditions can be obtained through our approach.

We start repeating the steps performed for the two previous maps to investigate how the
distance between $x$ and $x'$ changes with $\tstar$ in Eq.~(\ref{eq:test_delta}).
The results shown in Fig.~\ref{fig:standardMap} show that, contrary to what was
observed in the  previous maps, trajectories remain close to $\vx_t$ at least up to a time $\tstar$, as designed in
original formulation~(\ref{eq:deltax}).  Comparing to the results in Fig.~\ref{fig:maneville_conditional_trajs} for the
Pomeau-Maneville map, there seems to be no special time $t_i$ at which trajectories behave
fundamentally different.

The result above indicates that $\deltax$ in Eq.~(\ref{eq:deltax}) can be used to control
the time up to which trajectories are close. Even without a theoretical result indicating
what $\tstar(\vx)$ should be used, we see from the results above that $\tstar = 0.5 t_e$
is an heuristic choice that guarantees $x'$ with $E_{x'}$ both higher and lower than $E_x$
for different values of $t_e(x)$. Implementing this into a Metropolis-Hastings sampling
method we systematically observe that the acceptance of our method approaches zero,
ruining the efficiency of our sampling algorithm. To
understand why this happens, recall that that a crucial simplification in the derivation of Eq.~(\ref{eq:energy_condition})
was that of reversible proposal distribution g(x'|x), Eq.~(\ref{eq.approx}).
It will be argued below that the proposal with $\tstar(\vx) = 0.5 t_e(\vx)$ in the
standard map guarantees a bounded $\pi(E_{\vx'})/\pi(E_\vx)$, but it fails to guarantee a
constant acceptance ratio, Eq.~(\ref{eq:acceptance}), because of the mismatch in the
proposal distributions $g(x'|x)$ and $g(x|x')$, in violation of assumption~(\ref{eq.approx}).
In fact, results in Fig.~\ref{fig:standardMap} suggest that, even though the escape
time for $\tstar = 0.5 t_e$ lead to similar $\log t_e$, their respective FTLE varies
dramatically.
For a local proposal drawn from a half normal distribution -- see
Appendix~\ref{sec.halfgaussian} --  the ratio of the proposals $g$ is given by 
\begin{equation}
\frac{g(\vx|\vx')}{g(\vx'|\vx)} = \frac{\deltax}{\delta_x(\vx')} \exp \left[ -\frac{\pi|\vx' - \vx|^2}{4 \deltax^2} \left(1 - \frac{\deltax^2}{\delta_x(\vx')^2} \right) \right]
\end{equation}
By definition, $\vx'$ is constructed to be drawn such that $|\vx' - \vx| \approx \deltax$.
Thus, the ratio $g(\vx|\vx')/g(\vx'|\vx)$ essentially depends on the ratio $R(\vx,\vx') \equiv \deltax/\delta_x(\vx')$.
This allows to write $g(\vx|\vx')/g(\vx'|\vx) = f(R(\vx,\vx'))$ as
\begin{equation}
f(R) = R \exp \left[ - \left(1 - R^2 \right)\pi/4 \right] \ \ .
\end{equation}
This function fulfills $f(1) = 1$ and $f(0) = 0$, and decreases to zero for $R \ll 1$ and $R \gg 1$.
Thus, the more the distributions differ, the larger/smaller $f(R)$ is (depending on whether $R<1$ or $R>1$).
To guarantee a constant acceptance, the proposal distribution also needs to guarantee a bounded ratio $g(\vx|\vx')/g(\vx'|\vx)$, which thus equates to guarantee a bounded $R(\vx',\vx)$.
Since there are no more free parameters of the proposal distribution, what remains to be analyzed is whether $R(\vx',\vx)$ is bounded or not.
Figure~\ref{fig:standardMap3} illustrates the different values of $R(\vx',\vx)$ obtained from the same points used to construct the histogram of Fig.~\ref{fig:standardMap}, for the case $t_e = 16437$.
As anticipated, it indicates, that the ratio $R(\vx',\vx)$ is orders of magnitude different from 1 specially with $t_e(\vx') > t_e(\vx)$, which, from the preceding discussion, leads to an arbitrarily small acceptance rate.

In summary, this section showed how the proposal distribution with $\tstar(\vx) = 0.5
t_e(\vx)$ in the open standard map allows to propose states with an increasing $\log
t_e(\vx')$. This can be used, e.g., for algorithms that aim to \emph{find} long living trajectories.
However, this proposal distribution leads to vanishing acceptance rate in a Metropolis-Hastings simulation, which implies that it is not suitable to \emph{sample} long living trajectories.
To understand the reason for the vanishing acceptance, consider a proposed move from $x$ to $x'$ that leads to the desired local increase of $E=\log t_e$. In order for this move to be accepted, as discussed in Sec.~\ref{sec.2}, the probability of the reverse move (from $x'$ to $x$) should not be vanishingly small (ideally, $g(x|x') \approx g(x'|x)$). However, a local move in $E=\log t_e$ is a large move in $t_e$ (for large $E$), which is multiplied by the FTLE $\lambda_{\tstar}(x)$ and exponentiated to compute the characteristic search scale $\deltax$ in Eq.~(\ref{eq:deltax}). In the standard map, the FTLE is not sufficiently small to compensate for this increase and therefore the value of $\delta_{x'}(x') \ll \deltax$, leading to $g(x|x') \ll g(x'|x)$. The difference to the Pomeau-Maneville map is that the stickiness in the standard map is weaker ($\alpha>2$, larger FTLE) and more complicated~\cite{Meiss1986,Cristadoro2008} (e.g., not a single trapping point).
A possible strategy to obtain an efficient sample is to include the ratio of the proposal
distributions on the acceptance rate, potentially leading to an extension of Eq.~(\ref{eq:energy_condition}) which sets a tighter condition for a bounded acceptance rate.

\section{Conclusions}

Sampling rare trajectories in chaotic systems invoslves coupling two dynamical
  systems: the deterministic system we aim to study and the stochastic sampling method we
  construct. In a Metropolis-Hastings Monte Carlo sampling, the efficiency of the sampling
  depends critically on how the coupling is set through the choice of the proposal
  distribution.  The ideas presented in this paper show how to construct such proposal distribution
  based on the properties of the deterministic system.  In particular, we discussed how
  weakly-chaotic properties of the dynamics affect  the proposal distribution under naive
  assumptions of strong chaos.

  The main computational problem we discussed was to sample trajectories with very large
  escape time in open (weakly-chaotic) systems.  For the case of one-dimensional maps with
  marginal points, we were able to obtain an efficient Metropolis-Hasting method to sample
  trajectories. For the case of area-preserving maps with mixed phase space, we showed how
  our approach is able to {\it find} long-living trajectories in the system but that the
  Metropolis-Hasting method fails due to a low-acceptance ratio. We hope these results
  will trigger further work on the application of Monte Carlo methods in deterministic dynamical systems, in particular to the open problem of having an efficient sampling method to
  estimate the tails of the survival probability in intermittent systems (e.g., Hamiltonian systems with mixed phase space in arbitrary dimensions).

\section{Acknowledgments}

The results of this paper were obtained while JCL was a PhD student~\cite{jorgeThesis} at the MPIPKS in Dresden, funded by
Erasmus Grant No. 29233-IC-1-2007-1-PT-ERASMUS-EUCX-1. EGA was funded by the University of Sydney bridging Grant G199768.

  \appendix

  \section{Half-Gaussian local proposal}\label{sec.halfgaussian}

Our local proposal consists in perturbing $\vx$ by a finite amount $\vdelta$, $\vx' = \vx + \vdelta$, characterized by a direction $\vddelta$ and a norm $\delta$, $\vdelta \equiv \vddelta \delta$, $g(\vx'|\vx) = \vx + P(\vdelta|\vx)$.
A common case is when the probability distribution is separated in two independent terms~\cite{RobertCasellaBook}:
\begin{equation}
P(\vdelta|\vx) = P(\hat{\delta} | \vx) P(\delta | \vx),
\end{equation}
$P(\hat{\delta} | \vx)$ is uniformly distributed in the $D$ directions, and $P(\delta | \vx)$ has zero mean (i.e. an isotropic proposal).
Additionally, we  consider that $P(\delta | \vx)$ is characterized by a well defined scale, e.g. a half-normal distribution with mean $\deltax$:
\begin{equation}
P(\delta | \vx) = \frac{\sqrt{2}}{\sqrt{\pi \deltax^2}} e^{- \frac{\pi \delta^2}{4\deltax^2}} \text{ for } \delta > 0 \ \ .
\label{eq:isotropic_proposal}
\end{equation}
The main motivation for this choice is that the proposal distribution is described by a single function, $\deltax$, that quantifies the distance $\vx'-\vx$, $\Exp{|\vx' - \vx| |\vx} = \deltax$.

\end{document}